\documentclass[aps,pra,twocolumn,a4paper,superscriptaddress,showpacs,halfspacing,10pt]{revtex4-1}

\usepackage{graphicx}
\usepackage{amsmath}
\usepackage{upgreek}
\usepackage{hhline}
\usepackage{gensymb}

\begin{document}
\title{Short-wavelength out-of-band EUV emission from Sn laser-produced plasma}
\author{F Torretti}\email{f.torretti@arcnl.nl}
\affiliation{Advanced Research Center for Nanolithography, Science Park~110, 1098~XG Amsterdam, The Netherlands}
\affiliation{Department of Physics and Astronomy, and LaserLaB, Vrije Universiteit, De Boelelaan 1081, 1081 HV Amsterdam, The Netherlands}
\author{R Schupp}
\affiliation{Advanced Research Center for Nanolithography, Science Park~110, 1098~XG Amsterdam, The Netherlands}
\author{D Kurilovich}
\affiliation{Advanced Research Center for Nanolithography, Science Park~110, 1098~XG Amsterdam, The Netherlands}
\affiliation{Department of Physics and Astronomy, and LaserLaB, Vrije Universiteit, De Boelelaan 1081, 1081 HV Amsterdam, The Netherlands}
\author{A Bayerle}
\affiliation{Advanced Research Center for Nanolithography, Science Park~110, 1098~XG Amsterdam, The Netherlands}
\author{\mbox{J Scheers}}
\author{W Ubachs}
\affiliation{Advanced Research Center for Nanolithography, Science Park~110, 1098~XG Amsterdam, The Netherlands}
\affiliation{Department of Physics and Astronomy, and LaserLaB, Vrije Universiteit, De Boelelaan 1081, 1081 HV Amsterdam, The Netherlands}
\author{R Hoekstra}
\affiliation{Advanced Research Center for Nanolithography, Science Park~110, 1098~XG Amsterdam, The Netherlands}
\affiliation{Zernike Institute for Advanced Materials, University of Groningen, Nijenborgh 4, 9747 AG Groningen, The Netherlands}
\author{O O Versolato}
\affiliation{Advanced Research Center for Nanolithography, Science Park~110, 1098~XG Amsterdam, The Netherlands}

\onecolumngrid
Accepted for publication: F~Torretti~\emph{et~al}, J. Phys. B, in press (2018) DOI: \href{https://doi.org/10.1088/1361-6455/aaa593}{10.1088/1361-6455/aaa593}

\begin{abstract}
We present the results of spectroscopic measurements in the extreme ultraviolet (EUV) regime (7--17\,nm) of molten tin microdroplets illuminated by a high-intensity 3-J, 60-ns Nd:YAG laser pulse. The strong 13.5\,nm emission from this laser-produced plasma is of relevance for next-generation nanolithography machines. Here, we focus on the shorter wavelength features between 7 and 12\,nm which have so far remained poorly investigated despite their diagnostic relevance. Using flexible atomic code calculations and local thermodynamic equilibrium arguments, we show that the line features in this region of the spectrum can be explained by transitions from high-lying configurations within the Sn$^{8+}$--Sn$^{15+}$ ions. The dominant transitions for all ions but Sn$^{8+}$ are found to be electric-dipole transitions towards the $n$=4 ground state from the core-excited configuration in which a 4$p$ electron is promoted to the 5$s$ sub-shell. Our results resolve some long-standing spectroscopic issues and provide reliable charge state identification for Sn laser-produced plasma, which could be employed as a useful tool for diagnostic purposes. \\

\end{abstract}

\maketitle

\section{Introduction}

Sn and its highly charged ions are of undoubtable technological importance, as these are the emitters of extreme ultraviolet (EUV) radiation around 13.5\,nm used in nanolithographic applications \cite{Benschop2008,Banine2011}. In such state-of-the-art lithography machines, EUV light is generated using pulsed, droplet-based, laser-produced plasma (LPP) \cite{Fomenkov2017,Kawasuji2017}. Molten Sn microdroplets are illuminated by high-intensity (10$^{9}$--10$^{12}$\,W\,cm$^{-2}$) laser pulses, generating typically high-density (10$^{19}$--10$^{21}$\,electrons\,cm$^{-3}$) plasma. In this plasma, laser light is converted efficiently into photons with wavelengths close to 13.5\,nm. Technologically this is advantageous as it corresponds to the peak reflectivity of the mirrors composing the projection optics in nanolithography machines. These molybdenum-silicon multi-layer mirrors (MLMs) \cite{Bajt2002,Huang2017} are characterised by an ``in-band'', 2-\% reflectivity bandwidth centred around 13.5\,nm, which conveniently overlaps with the strong EUV emission of Sn LPP. The relatively high conversion efficiency of laser into EUV light in this wavelength region is mainly due to the atomic structure of the highly charged Sn ions found in this laser-produced plasma. 

The atomic transitions responsible for these EUV photons are 4$p^6$4$d^m$--4$p^6$4$d^{m-1}$4$f$ + 4$p^6$4$d^{m-1}$5$p$ + 4$p^5$4$d^{m+1}$, with $m$=6--0 for Sn$^{8+}$--Sn$^{14+}$ \cite{OSullivan2015}. These transitions are clustered in so-called unresolved transition arrays (UTAs) \cite{Bauche1988transition} as the close-lying large number of possible transitions arising from the complex open-4$d$-subshell electronic structure renders them unresolvable in practical applications. Configuration-interaction between the excited states 4$p^6$4$d^{m-1}$4$f^1$ and 4$p^5$4$d^{m+1}$ causes a significant redistribution of oscillator strength towards the high-energy side of the transition arrays, which is referred to as ``spectral narrowing'' \cite{hotplasmas}. A serendipitous level crossing involving the EUV-contributing excited configurations \cite{Windberger2015} furthermore fixes the average excitation energies of the excited states to the same value across a number of charge states.
\begin{figure*}
\includegraphics[scale=1]{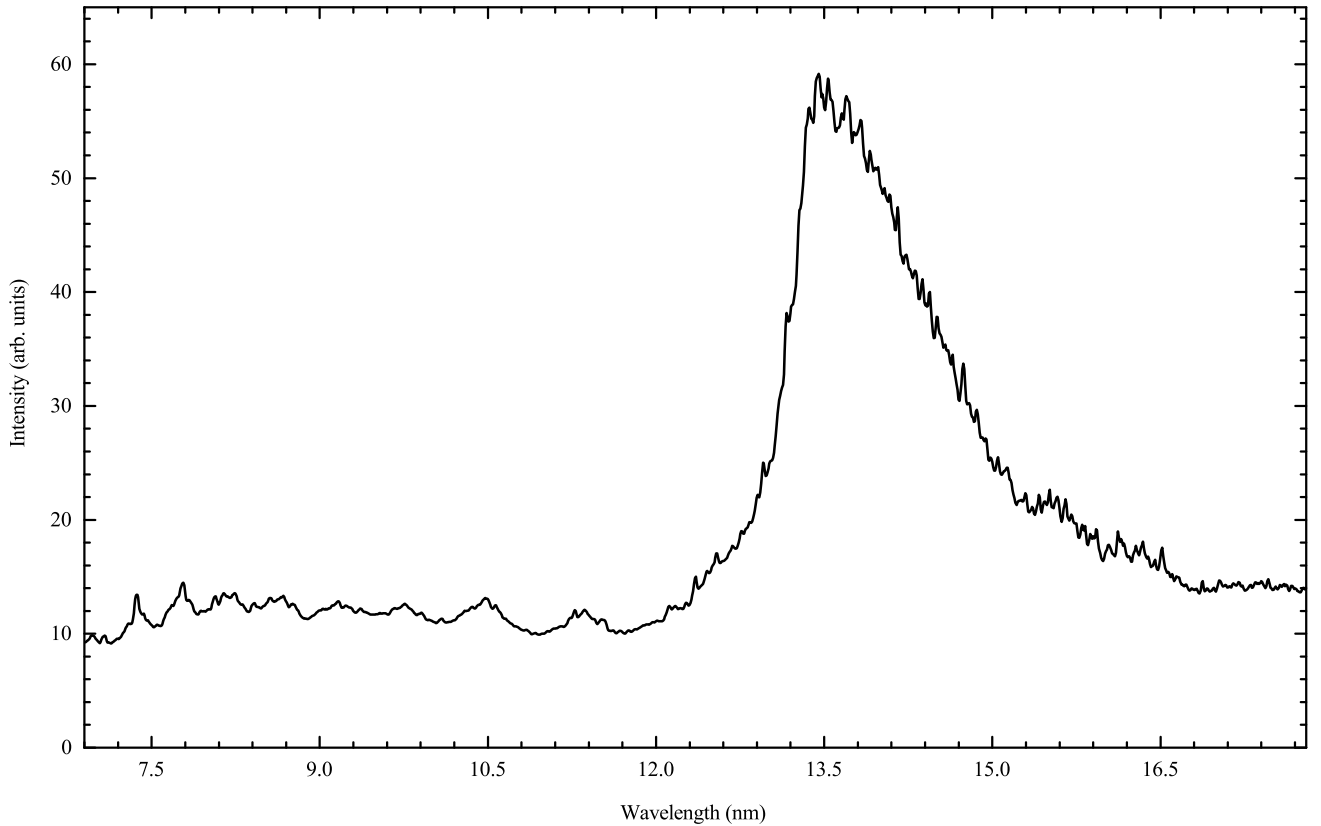}
\caption{Experimental spectrum emitted by the plasma generated from a 45\,$\upmu$m-diameter Sn droplet irradiated by a 60-ns Nd:YAG laser pulse, circular flat-top beam spot of 200\,$\upmu$m ($1/e^2$ encircled energy). This spectrum is obtained at an average laser intensity of $1.6 \cdot 10^{11}$\,W\,cm$^{-2}$.
\label{fig:1}}
\end{figure*}%

The nature of the convoluted structure of these highly charged ions has been addressed in several theoretical and experimental investigations both directly in the EUV regime and in the optical regime where charge-state-resolved spectroscopy enables complementary investigations that challenge the direct EUV measurements \cite{Azarov1993,Svendsen1994,Churilov2006SnIX--SnXII,Churilov2006SnVIII,Churilov2006SnXIII--XV,Ryabtsev2008SnXIV,Tolstikhina2006ATOMICDATA,DArcy2009a,Ohashi2010,Windberger2016,Colgan2017,Torretti2017}. 

In view of the application perspective, most of the work has so far been focused on the configurations responsible for emission around 13.5\,nm. However, in dense and hot plasmas, a significant amount of energy can be radiated at shorter wavelengths arising from configurations that remain poorly investigated with, to the best of our knowledge, but a single study \cite{Svendsen1994} dedicated to the corresponding transition arrays. This short-wavelength, ``out-of-band'' radiation could very well negatively affect the optics lifetime \cite{Huang2017,Mertens2004}, whilst obviously hindering the conversion efficiency from laser energy to 13.5\,nm photons.

In our experiments EUV spectra are obtained from plasma created by irradiating micrometer-sized molten Sn droplets with a pulsed Nd:YAG laser. We focus on the short-wavelength, high-energy features in the 7--12\,nm region and provide a detailed study of their origins using the flexible atomic code (\textsc{FAC}) \cite{Gu2008}. Applying a simple local thermodynamic equilibrium (LTE) scaling argument, we show that the line features in the experimental spectrum can be well explained by electric dipole ($E$1) transitions from high-lying configurations within the same tin charge states responsible for 13.5\,nm radiation in EUV sources. We find that the dominant contribution in the short wavelength region actually comes from the core-excited configuration 4$p^5$4$d^{m}$5$s$ where a 4$p$ core electron is promoted to the 5$s$ subshell. Our calculations excellently reproduce the line emission features observed experimentally, furthering the understanding of these emission features, and enabling the identification of individual contributions from various charge states Sn$^{8+}$--Sn$^{15+}$ which remain unresolvable in the 13.5\,nm UTAs. Furthermore, using the Bauche-Arnoult UTA formalism \cite{Bauche1988transition,hotplasmas,Bauche1984} as well as Gaussian fits to our results, we provide simplified outcome of our calculations which can be used to straightforwardly interpret Sn LPP spectra. 

\section{Experiment}

The experimental setup to generate droplets has been described in a previous publication\,\cite{Kurilovich2016}, and only the details relevant to this article are presented in the following. Droplets of molten tin of 99.995\,\% purity are dispensed from a droplet generator operating at a 10.2-kHz repetition rate inside a vacuum vessel filled with a continuous flow of Ar buffer gas ($\sim 10^{-2}$\,mbar). The droplets have a diameter of approximately 45\,$\upmu$m. An injection-seeded, 10-Hz repetition rate, 3-J, 60-ns pulse length Nd:YAG laser operating at its fundamental wavelength $\lambda$=1064\,nm is imaged to a circular flat-top beam spot (200\,$\upmu$m at $1/e^2$ encircled energy) at the droplet position, producing an averaged intensity of $1.6 \cdot 10^{11}$\,W\,cm$^{-2}$, close to industrially relevant conditions for obtaining high conversion efficiency \cite{George2007,Tanaka2005,Basko2016}. Using polarizing optics the laser energy can be set without modifying the beam spot. The linearly-polarised laser pulse is timed to hit the droplet, thus creating an EUV-emitting plasma. This emitted EUV radiation is coupled into a grazing-incidence spectrometer positioned at a 120\degree{} angle with respect to the laser light propagation direction (directly facing the laser-droplet interaction zone). Optical light and debris from the laser-produced plasma are blocked by a 200-nm-thick zirconium filter. In the spectrometer the EUV light is diffracted by a gold-coated concave grating (1.5\,m radius of curvature, 1200 lines-per-mm) at an angle of incidence of 86\degree{} with respect to the grating normal. The 100-$\upmu$m entrance slit and the detector, a Greateyes back-illuminated charge coupled device (\textsc{GE\,2048\,512\,BI\,UV1}) cooled to 0\degree C, are positioned on the circle determined by the radius of curvature of the grating, in a standard Rowland circle geometry. Wavelength calibration is performed after the measurements using an aluminium solid target positioned at the same location as the droplets and ablated by the Nd:YAG laser. Well-known Al$^{3+}$ and Al$^{4+}$ lines in the 10--16\,nm range from the NIST atomic database \cite{NIST_ASD} are used, obtaining a calibration function with a systematic one-standard-deviation uncertainty of 0.003\,nm. The typical full-width at half-maximum of the features observed is approximately 0.06\,nm at 13\,nm wavelength.

During acquisition, the camera is exposed to about ten droplets irradiated by the Nd:YAG laser. This exposure is repeated multiple times to improve the statistics of the measurement, resulting in an averaged spectrum. The obtained spectrum is corrected for geometrical aberrations causing the spectra to move across the vertical, non-dispersive axis of the camera, as is known from the Al calibration spectra. Background counts are dominated by readout-noise and are subtracted before averaging over the non-dispersive axis. The contribution of second-order diffraction is obtained from the line emission in the Al calibration spectra, enabling the related correction. Further corrections are subsequently applied for the sensor quantum efficiency, as obtained from the manufacturer datasheets, and the Zr-filter transmission curve \cite{Henke1993}. No correction curves could be obtained for the gold-coated grating. However, the behaviour in the 7--12\,nm region can reasonably be expected to be rather constant \cite{Henke1993}.

Figure \ref{fig:1} shows a thus obtained spectrum. The strongest feature is found near 13.5\,nm wavelength as expected. The left shoulder of this UTA feature, starting its steep rise upward from about 13\,nm with well-known contributions from Sn$^{10+}$--Sn$^{14+}$ \cite{Churilov2006SnIX--SnXII,Churilov2006SnXIII--XV,Ryabtsev2008SnXIV} continuing to a peak at exactly 13.5\,nm after which the feature decays more slowly as it moves over the contributions of the lower tin charge states \cite{Churilov2006SnIX--SnXII,Churilov2006SnVIII} whilst also suffering opacity-related broadening \cite{Fujioka2005}. This main feature appears to lie on top of an apparent continuum which extends over the full observed wavelength range. At the typical plasma conditions of this LPP, such continuum radiation is usually attributed to recombination processes \cite{Sullivan2009}. Continuing our investigations to line features with wavelengths below 12\,nm first requires new calculations since, as pointed out in the introduction, pertinent experimental data and calculations are sparsely available.

\section{Calculations}\label{sec:theory}

\subsection{Atomic structure}
For the interpretation of the short-wavelength side of the obtained experimental spectrum, we employ the flexible atomic code (\textsc{FAC}) \cite{Gu2008}. Specifically, we use it to investigate short-wavelength transitions in the 7--12\,nm region in Sn$^{8+}$--Sn$^{15+}$. \textsc{FAC} performs relativistic atomic structure calculations including configuration-interaction. The atomic wavefunctions are calculated as linear combination of configuration state functions, which are determined from a local central potential obtained by solving self-consistently the Dirac equations. Relativistic effects are taken into account by the Dirac-Coulomb Hamiltonian. Radiative transition rates are calculated from the obtained wavefunctions in the single-multipole approximation. For more details, we refer to reference \cite{Gu2008}.

Calculations are performed including the following configurations for the open-4$d$-shell ions Sn$^{8+}$--Sn$^{13+}$: the ground state [Kr]4$d^m$, [Kr]4$d^{m-1}$4$f$, [Kr]4$d^{m-1}$5$pf$, [Kr]4$d^{m-1}$6$pf$, [Ar]3$d^{10}$4$s^2$4$p^5$4$d^{m+1}$, [Ar]3$d^{10}$4$s^2$4$p^5$4$d^{m}$5$sd$, [Ar]3$d^{10}$4$s^2$4$p^5$4$d^{m}$6$sd$ ($m$=6--1). This somewhat limited set of configurations is used since, in the chosen benchmark case of Sn$^{8+}$, they give good agreement with both \emph{ab initio} multi-configuration Dirac-Fock calculations performed by Svendsen and O'Sullivan \cite{Svendsen1994} as well as with the current experimental observations. For the open-4$p$-shell ions Sn$^{14+}$ and Sn$^{15+}$, the configuration sets used are [Ar]3$d^{10}$4$s^2$4$p^{q}$, [Ar]3$d^{10}$4$s^2$4$p^{q}$4$df$, [Ar]3$d^{10}$4$s^2$4$p^{q}$5$spd$, [Ar]3$d^{10}$4$s^2$ 4$p^{q}$6$spd$ ($q$=4,5), which are found to give good agreement with measured features. Weighted transition rates $g_i A_{ij}$ (statistical weight $g_i$ of upper level $i$ times the transition probability $A_{ij}$ from $i$ to $j$) are calculated for transitions from each excited state towards the ground state for all ions involved. 
\begin{figure}
\includegraphics[scale=1]{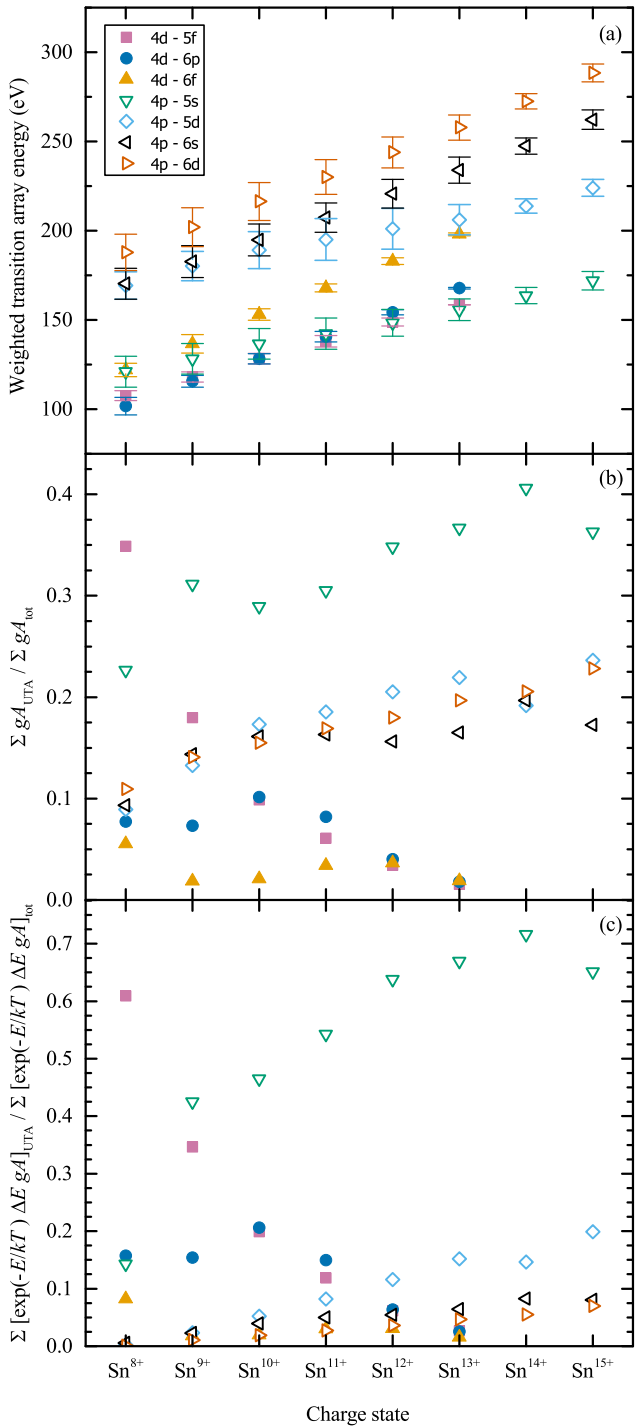}
\caption{Scaling of the properties of the investigated transition arrays along the isonuclear sequence of the Sn ions observed in the experimental spectrum. The legend denotes the transition arrays according to the nomenclature used in the main text. (a) Weighted transition array energies $\upmu_1$ (see section \ref{sec:UTA}) with the error bars indicating the width of the arrays, expressed as the Gaussian standard deviation $\sigma$. (b) Relative contributions of transition arrays for each ion in terms of weighted transition rates $gA$. (c) Relative array intensities for each ion calculated assuming the excited states relative populations to be in accordance with local thermodynamic equilibrium scaling (see main text).
\label{fig:2}}
\end{figure}%

\subsection{Emission properties}
Our \textsc{FAC} calculations indicate that in the 7--12\,nm (or approximately~110--180\,eV) region the main transitions of interest (i.e.~the strongest contributions to the radiative decay) are the following: 4$d^m$--4$d^{m-1}$5$f$ (from here on denoted as 4$d$--5$f$), 4$d^m$--4$d^{m-1}$6$p$ (4$d$--6$p$), 4$d^m$--4$d^{m-1}$6$f$ (4$d$--6$f$), 4$p^6$4$d^m$--4$p^5$4$d^{m}$5$s$ (4$p$--5$s$), 4$p^6$4$d^m$--4$p^5$4$d^{m}$5$d$ (4$p$--5$d$), 4$p^6$4$d^m$--4$p^5$4$d^{m}$6$s$ (4$p$--6$s$), 4$p^6$4$d^m$--4$p^5$4$d^{m}$6$d$ (4$p$--6$d$). These are all $E$1 single-electron excitations. In figure \ref{fig:2} the properties of these transition arrays are presented for the isonuclear sequence Sn$^{8+}$--Sn$^{15+}$. The weighted mean energies in figure \ref{fig:2}(a) are calculated using as weights the transition rates (this is in accordance with the UTA formalism, see section \ref{sec:UTA} for a more thorough description). They are seen to scale quite regularly along the isonuclear sequence, increasing for higher ionic charge. There is no level crossing apparent, making these transition arrays a potential diagnostic tool to identify the contributions of different charge states in the spectra of Sn LPPs. 

The weighted mean energies of the transition arrays towards the $n$=4 ground state stemming from configurations with principal quantum number $n$=6 are observed to scale more strongly with charge state $Z$ compared to configurations having $n$=5. This can be explained similarly to hydrogenic scaling of binding energy for different $n$ orbitals. In these systems, the binding energy scales with $Z^2$\,$n^{-2}$. The transition energy, i.e.~the difference between the ground state and the excited state binding energies, will therefore have a stronger dependence on ionic charge for higher $n$. Naturally, this exact scaling does not perfectly describe the one observed, as hydrogenic approximations are far too limited to be able to interpret complicated ionic systems with multiple valance electrons such as the highly charged Sn ions here considered.
\begin{figure*}
\includegraphics[scale=1]{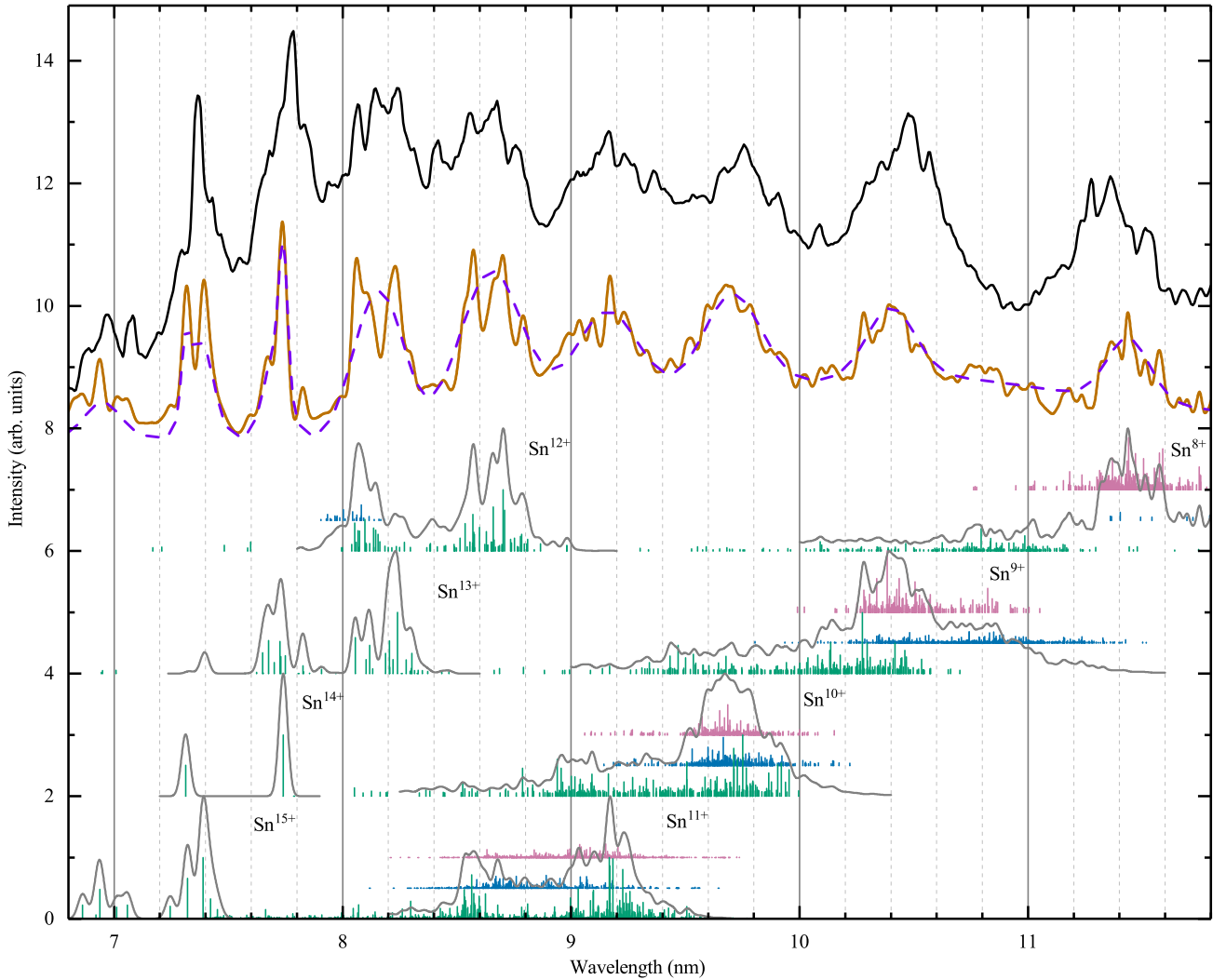}
\caption{The topmost black line shows the short wavelength features from the experimental spectrum in figure \ref{fig:1} in the 7--12\,nm range. The bar plots show the LTE-weighted transition rates for the three strongest transition arrays of each ion (see main text and figure \ref{fig:2}): top, in pink, 4$p^6$4$d^m$--4$p^6$4$d^{m-1}$5$f$; middle, in blue, 4$p^6$4$d^m$--4$p^6$4$d^{m-1}$6$p$; bottom, in green, 4$p^6$4$d^m$--4$p^5$4$d^{m}$5$s$. For clarity, only the top 95~\% of the transition rates is shown. Solid grey lines represent the envelopes for each charge state, determined by the convolution of all the transition arrays (see main text) with Gaussian curves. The sum of all the normalised envelopes, scaled with a single multiplication factor and shifted upwards with an offset for improved visibility, is shown by the dark-orange line below the experimental data. The purple dashed line is the result of using the spectral shape parameters (table \ref{tab:2}) introduced in section \ref{sec:spectra_shape_parameters}.
\label{fig:3}}
\end{figure*}%

For each ion figure \ref{fig:2}(b) shows the relative contributions of the arrays using the sum of their UTA transition rates $\Sigma gA_{UTA}$ relative to the total $\Sigma gA_{tot}$ for all the UTAs here investigated of that charge state. The transition array with the highest relative contribution is the 4$p$--5$s$ array with the sole exception of Sn$^{8+}$, for which the largest contribution comes from the 4$d$--5$f$ array. Relatively large weighted transition rates are seen also for the other core-excited transitions 4$p$--5$d$, 4$p$--6$s$, and 4$p$--6$d$. In previous work \cite{Svendsen1994} only the arrays 4$d$--5$f$ and 4$p$--6$d$ were taken into account to explain their experimental result. These configurations, according to our \textsc{FAC} calculations, are of significance primarily for the ions Sn$^{8+}$--Sn$^{11+}$ but cannot be expected to sufficiently explain the features even for these charge states. For all the ions investigated, using the weighted transition rates we show that the core-excited, open-4$p$-shell configurations play a significant role in interpreting the high-energy out-of-band EUV emission of these Sn ions. We note that the energies of the associated configurations scale steeply with ionic charge. This means that we have to take into account the fact that the transition energies quickly move over the wavelength band studied in this work, as well as that the relative populations of the excited states need to be considered carefully.

\subsection{LTE considerations}
To obtain the actual expected emission characteristics in a plasma, weighted transition rates of the relevant UTAs alone are not sufficient, and the relative population of the excited states needs to be considered. A straightforward approach is provided by the local thermodynamic equilibrium (LTE) assumption, the conditions for which are expected to be well met in the extremely high-density plasma created by the Nd:YAG laser pulse \cite{Colombant1973} where collisional processes outpace the relevant atomic decay rates by orders of magnitude. In this case, the population of the excited states can be approximated by the related Maxwell-Boltzmann statistics. The intensity $I_{ij}$ of a single transition between atomic states $i$ and $j$ can subsequently be expressed as \cite{hotplasmas}
\begin{equation}
I_{ij} = n_i \Delta E_{ij} A_{ij} \propto \exp (-E_i/kT) \Delta E_{ij} g_i A_{ij}, \label{eq:MB_LTE}
\end{equation}
where $n_i$ is the population of the excited state, $\Delta E_{ij}$ is the transition energy, $E_i$ is the energy of the excited state and $kT$ is the temperature associated with the emitter. From a preliminary comparison between calculations and the experimental spectrum, it becomes apparent that the wide wavelength range of the line emission is characterised by a very broad charge state distribution: in the spectrum, features belonging to Sn$^{8+}$ through Sn$^{15+}$ are tentatively observed. This broad distribution hints to the fact that the spectrum cannot be modelled by a single temperature, and that the different charge states exist in regions of the plasma where the temperature and density allow them to exist in the first place. Thus, in the following, the temperature $kT$ in equation \eqref{eq:MB_LTE} is assumed to be the temperature at which the average charge state in the plasma is equal to the ion under consideration. The required relation between average charge state $\bar{Z}$ and temperature $kT$ is determined from thermodynamic considerations of the equation of state of Sn in reference \cite{Basko2015}, obtaining the following relationship:
\begin{equation}
kT \mathrm{(eV)} = 0.56 \cdot \bar{Z} ^{5/3} . \label{eq:kT}
\end{equation}
As is shown in figure \ref{fig:2}(c), once this scaling is applied, the 4$p$--5$s$ re-enforces its position as dominant contribution to the ions' emitted radiation. The arrays 4$d$--5$f$ and 4$d$--6$p$ are relatively bright for the lower charge states Sn$^{8+}$--Sn$^{11+}$, whilst the core-excited transitions become more relevant for the higher charge states.

\section{Comparison}\label{sec:comparison}

Figure \ref{fig:3} shows the comparison between resulting LTE-weighted \textsc{FAC} calculations and the experimental spectrum. The former contribution is broken down into its individual charge state constituents for which we show drop-line plots of the three largest contributing configurations in each case. Only the top 95\% of the transition rates is shown for improved visibility. The envelopes shown, again per charge state, are the results of convolving the quantities $\Delta E_{ij} \exp (-E_{i}/kT) g_i A_{ij}$ for all contributing configurations, with a Gaussian function having width equal to the instrument resolution. The sum of all the normalised envelopes without any further weighting factors is shown as the solid trace (dark-orange) below the experimental spectrum. This trace is only arbitrarily scaled with a common, unity multiplication factor and shifted upwards using a constant offset for improved visibility without the use of any free fit-parameters.  Nearly all sub-structures visible within the experimental spectrum can be linked to their theoretical counterparts. We note that this treatment treatment does not include opacity effects. This is justified by previous experimental and theoretical studies of tin plasmas \cite{Colgan2017,Fujioka2005}, where it was shown that opacity does not play a significant role in the 7--12 nm region, quite unlike the case of the 12--15 nm region spanned by the UTAs relevant for the in-band emission. The agreement of our calculations with the experiment supports this statement. 

Remarkably, our equivalent summation of the various normalised contributions reproduces the experimental spectrum quite well. This observation hints at a broad and relatively flat charge state distribution, which in part may be due to the time- and space-integrating nature of our measurement, thus averaging over large temperature and density gradients that are characteristic for these laser-produced plasmas \cite{Su2017}. The detailed explanation of this observed feature is left for future work, as enabled by the here presented line identification. Minor wavelength shifts between calculated features and measurement are apparent, with an average absolute difference of 0.04\,nm. Such a difference on the parts-per-thousand level is excellent considering the limited number of configurations used in our calculations, and surpasses the absolute accuracy obtained with FAC in complex ions where configuration interaction is prominent \cite{Kilbane2010}. We conclude that the line features in our final synthetic spectrum are in excellent agreement with the experimental spectrum, strongly supporting our identifications of both charge states and electronic configurations.

\section{UTA formalism}\label{sec:UTA}

Having obtained an excellent replication of the experiment using our calculations, we will next employ the Bauche-Arnoult UTA formalism \cite{Bauche1988transition,Bauche1984,hotplasmas} to interpret our calculation results. The transition arrays presented in the previous sections are here characterised using the moments of their distribution \cite{Bauche1988transition,hotplasmas}. The $n$th-order moment $\upmu_n$ of the energy distribution of $E$1 transitions between configurations $A$ and $B$ is expressed as
\begin{multline}
\upmu_n \left( A \rightarrow B \right) = \\ \frac{ \sum_{i,j} \left| \left< i \left| \textbf{D} \right| j \right> \right|^2 \left[ \left< i \left| \textbf{H} \right| i \right> - \left< j \left| \textbf{H} \right| j \right> \right]^n}{\sum_{i,j} \left| \left< i \left| \textbf{D} \right| j \right> \right|^2},
\end{multline}
where $\textbf{D}$ is the electric dipole operator, $\textbf{H}$ the Hamiltonian matrix, $i$ and $j$ denotes the atomic states belonging to configurations $A$ and $B$, respectively. The moment $\upmu_n$ can also be conveniently expressed in terms of transition energies and weighted transition rates \cite{White2005}:
\begin{equation}
\upmu_n \left( A \rightarrow B \right) = \frac{\sum_k (\Delta E_k)^n g A_k}{\sum_k g A_k},
\end{equation}
with $\Delta E_k$ the energy of the $k$th $E$1 transition (between atomic states $i$ and $j$), $g$ the statistical weight of the transition upper level and $A_k$ the Einstein's coefficient.

The first moment of the distribution, $\upmu_1$, is the $E$1-strength-weighted average energy of the transition array. The width of the transition array is represented by the standard deviation of the distribution, i.e.~the square root of the variance:
\begin{equation}
\sigma = \sqrt{V} = \sqrt{ \upmu_2 - (\upmu_1)^2}.
\end{equation}
Finally, the asymmetry of the distribution is described by the skewness $\alpha_3$ \cite{Bauche1984}:
\begin{equation}
\alpha_3 = \frac{\upmu_3 - 3 \upmu_1 V - (\upmu_1)^3}{V^{3/2}}.
\end{equation}

These quantities are helpful in the statistical representation of the transition array using a single skewed Gaussian curve \cite{Bauche1984}:
\begin{multline}
f(E) = \left\{ 1 - \frac{\alpha_3}{2} \left(\frac{E-\upmu_1}{\sigma} - \frac{(E-\upmu_1)^3}{3 \sigma^3} \right) \right\} \\ \exp \left[ - \frac{(E-\upmu_1)^2}{2 \sigma^2} \right].
\label{eq:sk_gauss}
\end{multline}
An example of the application of this representation is given in figure \ref{fig:4} for the three main transition arrays here studied of Sn$^{10+}$. It illustrates well the huge simplification that is still able to capture the emission characteristics using but three parameters, which are listed for all studied transition arrays and for all investigated Sn ions in Table \ref{tab:1}. The suitability of this approach in replicating the total ion emission, as shown in figure \ref{fig:4} by the solid grey and dotted orange curves, is discussed in the following section.

\section{Spectral shape parameters}\label{sec:spectra_shape_parameters}
\begin{figure}[t]
	\includegraphics[scale=1]{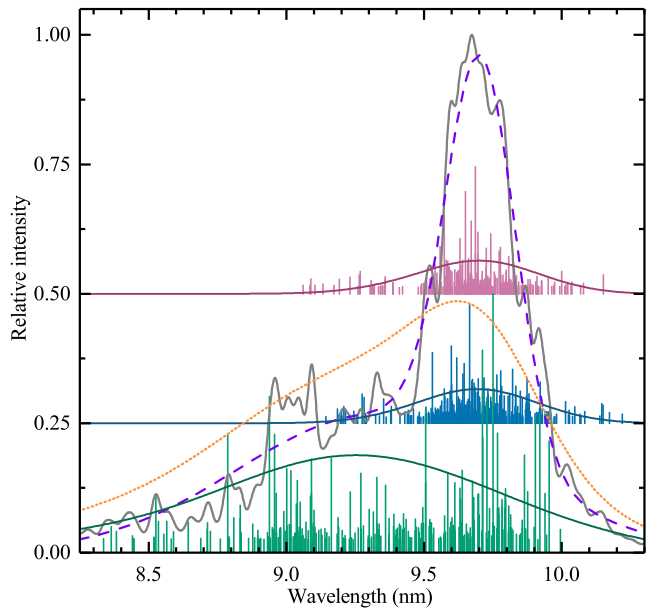}
	\caption{Details of the emission of Sn$^{10+}$ in the 8.2--10.4\,nm region. The Gaussian envelope of the emission (solid gray line, from figure \ref{fig:3}) is mainly determined by the LTE-weighted transition rates of the three strongest transition arrays, here shown as the bar plots: in pink (top) 4$p^6$4$d^4$--4$p^6$4$d^3$5$f$, in blue (middle) 4$p^6$4$d^4$--4$p^6$4$d^3$6$p$, and in green (bottom) 4$p^6$4$d^4$--4$p^5$4$d^{4}$5$s$. Only the top 95~\% of the transition rates is shown. Each transition array is also represented using their respective skewed Gaussian from equation \ref{eq:sk_gauss} with the relevant parameters from table \ref{tab:1}, scaled accordingly to their relative $gA$ contribution. Also shown is the total sum of the skewed Gaussians (dotted orange curve). The double Gaussian fit (purple dashed line) is determined by the spectral shape parameters in table \ref{tab:2}. All the curves have been scaled to conserve the integral emission of the transition arrays.
		\label{fig:4}}
\end{figure}%
The statistical quantities in the UTA formalism are useful to present in a concise manner the result of our \textsc{FAC} calculations, thus providing information about the atomic physics aspects behind the emission of these Sn ions. Whilst these values could be used to interpret experimental data, it is necessary to take into account the relevant configuration arrays by scaling their relative contributions accordingly to our LTE arguments (see figure \ref{fig:2}(c)). Even doing so, the UTA formalism would provide limited detail in the comparison with experimentally measured emission intensities as is illustrated in figure \ref{fig:4}. We instead provide a more straightforward and more apt comparison which can be used to identify the individual contributions of the Sn$^{8+}$--Sn$^{15+}$ charge states. We use the LTE-weighted emission intensities obtained convolving all the investigated transition arrays, i.e.~the Gaussian envelopes shown in figure \ref{fig:3}. These envelopes can be conveniently described by fitting two Gaussian curves, obtaining the very good qualitative agreement shown in figure \ref{fig:4}. The equation for these Gaussian fits reads
\begin{equation}
f(\lambda) = \Sigma_i A_i \exp \left[ -\frac{(\lambda - \lambda_{c,i})^2}{2 w_i^2} \right],\label{eq:gauss_fit}
\end{equation}
with $\lambda$ being the wavelength, and $i$=1,2. The list of coefficients $A_i$, $\lambda_{c,i}$, $w_i$ is given in table \ref{tab:2}. No simple scaling of these parameters with temperature is obtainable. Therefore, to extend this approach to arbitrary $kT$ values, our calculations need to be re-evaluated starting from equation \eqref{eq:MB_LTE}.

Despite their artificial meaning, these parameters can be very useful to easily diagnose the emission of Sn LPPs in this wavelength region. Whilst the substructure of the emission is lost the major features and characteristics are preserved, with much more detail and accuracy than in the skewed Gaussian representation using the UTA formalism. Figure \ref{fig:4} shows this aspect, with the skewed Gaussian curve overestimating the contribution in the short wavelength region, thus shifting the centre of mass of the emission to shorter wavelength. The data as provided is self-consistent, in the sense that the relative amplitudes of the Gaussian fits agree with the relative contribution to the spectrum of the different ions. Potentially, combining this tool with absolutely calibrated spectral measurement could yield valuable information regarding the relative charge state populations in the plasma, and the magnitude and nature of the radiation continuum underlying the atomic line emission. These aspects are topics of future investigations.

\section{Conclusions}

We present the results of spectroscopic measurements in the extreme ultraviolet regime of the emission of the plasma created from molten Sn droplets when irradiated by a high-energy pulse from a Nd:YAG laser at its fundamental wavelength. Using the flexible atomic code, electric dipole transition from excited configurations towards the ground states in the ions Sn$^{8+}$--Sn$^{15+}$ were investigated. Including a simple local thermodynamic equilibrium scaling for the relative populations of these excited states, we have shown that the most intense contribution to the radiation in the 7--12\,nm region can be attributed to the radiative decay of the core excited configuration [Ar]3$d^{10}$4$s^2$4$p^5$4$d^{m}$5$s$ towards the ground state in the ions here studied. Moreover, the transition energies and rates from \textsc{FAC} calculations, scaled with LTE population, reproduce very well the unresolved transition arrays measured experimentally. The unresolved-transition-array formalism is subsequently used to present in a concise manner the result of our atomic structure calculations. Furthermore, spectral shape parameters of double Gaussian fits to the LTE-weighted emission spectra of each Sn ion are provided,  enabling straightforward interpretation of our results. These parameters also provide simplified spectral information to interpret emission from industrial plasma EUV light sources which may facilitate their optimization. Our findings thus further the understanding of the atomic structure of Sn ions and their emissions in the context of laser-produced plasma and EUV sources.

\begin{acknowledgments}
This work has been carried out at the Advanced Research Center for Nanolithography (ARCNL), a public-private partnership of the University of Amsterdam (UvA), the Vrije Universiteit Amsterdam (VU), the Netherlands Organisation for Scientific Research (NWO) and the semiconductor equipment manufacturer ASML.

We would like to thank J.C.~Berengut (University of New South Wales, Sydney) for the fruitful conversation during the early stages of the manuscript. We would also like to thank T.A.~Cohen Stuart, R.~Jaarsma, the AMOLF mechanical and design workshop, the software and electronics departments for technical support.
\end{acknowledgments} 	

\setlength{\tabcolsep}{4.1pt}
\begin{table*}[]
\centering
\caption{UTA properties for all the $E$1-contributing transition arrays investigated with the flexible atomic code for the ions Sn$^{8+}$ to Sn$^{15+}$. Weighted mean energy $\upmu_1$, standard deviation $\sigma$, skewness $\alpha_3$ and total number of lines $N$ are here presented in accordance with the UTA formalism \cite{Bauche1988transition} (see section \ref{sec:UTA}). \smallskip}
\label{tab:1}
\begin{ruledtabular}
\begin{tabular}{crrrrcrrrrcrrrr}
 &\multicolumn{4}{c}{Sn$^{8+}$}& \rule[4.65mm]{0pt}{0mm}\rule[-1.8mm]{0pt}{0mm} &\multicolumn{4}{c}{Sn$^{9+}$}& &\multicolumn{4}{c}{Sn$^{10+}$}     \\\cline{2-5}\cline{7-10}\cline{12-15}
\multicolumn{1}{c}{UTA} & \rule[4.65mm]{0pt}{0mm}\rule[-2.3mm]{0pt}{0mm} $\upmu_1$\,(eV) & \multicolumn{1}{c}{$\sigma$\,(eV)} & \multicolumn{1}{c}{$\alpha_3$} & \multicolumn{1}{c}{$N$}  & & $\upmu_1$\,(eV) & \multicolumn{1}{c}{$\sigma$\,(eV)} & \multicolumn{1}{c}{$\alpha_3$} & \multicolumn{1}{c}{$N$} &  & $\upmu_1$\,(eV) & \multicolumn{1}{c}{$\sigma$\,(eV)} & \multicolumn{1}{c}{$\alpha_3$} & \multicolumn{1}{c}{$N$}   \\
4$d$--5$f$ & 107.6         & 2.7          & -0.520     & 5034&  & 118.0         & 2.9          & -0.214     & 5313 & & 128.3         & 3.0          & 0.323      & 3247  \\
4$d$--6$p$ & 101.7         & 4.9          & -0.576     & 3226  & & 115.7         & 3.3          & 0.090      & 3366&  & 128.2         & 2.9          & 0.209      & 2275  \\
4$d$--6$f$ & 122.0         & 3.7          & 0.310      & 5614&  & 136.6         & 5.2          & 0.091      & 5423&  & 153.0         & 3.3          & -1.077     & 2672  \\
4$p$--5$s$ & 121.0         & 8.6          & 0.779      & 4752  & & 128.2         & 8.6          & 0.869      & 6304 & & 136.6         & 8.6          & 0.746      & 3897  \\
4$p$--5$d$ & 169.3         & 7.7          & -0.541     & 20806 & & 180.2         & 8.2          & -0.530     & 28983&  & 189.1         & 10.4         & -0.487     & 21257 \\
4$p$--6$s$ & 170.3         & 8.7          & 0.741      & 6192 & & 182.7         & 8.9          & 0.797      & 8288&  & 194.8         & 8.9          & 1.018      & 5664  \\
4$p$--6$d$ & 187.9         & 10.2         & 0.418      & 21424& & 202.0         & 10.9         & 0.449      & 29012 & & 216.4         & 10.6         & 0.584      & \rule[-3.8mm]{0pt}{0mm}21331 \\
 &\multicolumn{4}{c}{Sn$^{11+}$}& \rule[-1.8mm]{0pt}{0mm} &\multicolumn{4}{c}{Sn$^{12+}$}& &\multicolumn{4}{c}{Sn$^{13+}$}     \\\cline{2-5}\cline{7-10}\cline{12-15}
\multicolumn{1}{c}{UTA} & \rule[4.65mm]{0pt}{0mm}\rule[-2.3mm]{0pt}{0mm} $\upmu_1$\,(eV) & \multicolumn{1}{c}{$\sigma$\,(eV)} & \multicolumn{1}{c}{$\alpha_3$} & \multicolumn{1}{c}{$N$} & & $\upmu_1$\,(eV) & \multicolumn{1}{c}{$\sigma$\,(eV)} & \multicolumn{1}{c}{$\alpha_3$} & \multicolumn{1}{c}{$N$}&  & $\upmu_1$\,(eV) & \multicolumn{1}{c}{$\sigma$\,(eV)} & \multicolumn{1}{c}{$\alpha_3$} & \multicolumn{1}{c}{$N$} \\
4$d$--5$f$ & 138.0         & 3.1          & 0.314      & 870 & & 148.8         & 2.2          & -1.061     & 96 &  & 158.4         & 0.1          & 10.042     & 3   \\
4$d$--6$p$ & 140.6         & 2.9          & -0.539     & 644 & & 154.3         & 1.5          & -1.369     & 64  & & 167.8         & 0.5          & 2.351      & 3   \\
4$d$--6$f$ & 167.9         & 2.2          & -0.606     & 679 & & 182.9         & 1.9          & -0.770     & 89 &  & 198.0         & 0.7          & 0.715      & 5   \\
4$p$--5$s$ & 142.3         & 8.8          & 0.953      & 1620 & & 148.4         & 7.5          & 1.371      & 377 & & 155.7         & 6.0          & 1.315      & 37  \\
4$p$--5$d$ & 195.0         & 11.7         & -0.025     & 8098 & & 201.0         & 11.4         & 0.486      & 1546 & & 206.2         & 8.5          & 1.373      & 132 \\
4$p$--6$s$ & 207.3         & 8.3          & 1.170      & 2112 & & 220.8         & 8.0          & 1.283      & 403  & & 233.9         & 7.3          & 1.471      & 38  \\
4$p$--6$d$ & 230.1         & 9.7          & 0.947      & 8126 & & 243.8         & 8.7          & 1.299      & 1564 & & 257.8         & 7.1          & 1.628      & \rule[-3.8mm]{0pt}{0mm}132\\
 &\multicolumn{4}{c}{Sn$^{14+}$}& \rule[-1.8mm]{0pt}{0mm} &\multicolumn{4}{c}{Sn$^{15+}$}& & & & & \\\cline{2-5}\cline{7-10}
\multicolumn{1}{c}{UTA} & \rule[4.65mm]{0pt}{0mm}\rule[-2.3mm]{0pt}{0mm} $\upmu_1$\,(eV) & \multicolumn{1}{c}{$\sigma$\,(eV)} & \multicolumn{1}{c}{$\alpha_3$} & \multicolumn{1}{c}{$N$} & & $\upmu_1$\,(eV) & \multicolumn{1}{c}{$\sigma$\,(eV)} & \multicolumn{1}{c}{$\alpha_3$} & \multicolumn{1}{c}{$N$} & & 	&	&	&	\\
4$p$--5$s$ & 163.6         & 4.5          & 0.541      & 3 & & 172.0         & 5.2 & 0.678      & 14 &  &  &  &  &  \\
4$p$--5$d$ & 213.8         & 4.0          & 1.037      & 7 & & 224.0         & 4.7          & 0.790      & 39  & &  &  &  &  \\
4$p$--6$s$ & 247.5         & 4.6          & 0.715      & 3 & & 262.2         & 5.4 & 0.744      & 14  &  &  &  &  \\
4$p$--6$d$ & 272.5         & 4.3          & 0.932      & 7 & & 288.4         & 5.1          & 0.750      & 40 & &  &  &\rule[-1.8mm]{0pt}{0mm}   & \\
\end{tabular}
\end{ruledtabular}
\end{table*} 
\newpage
\mbox{}
\newpage
\mbox{}
\begin{table*}[]
\centering
\caption{Coefficients of the Gaussian fits to the envelopes of the LTE-weighted intensities for all ions investigated (see figure \ref{fig:3} and section \ref{sec:comparison}) evaluated at the temperature $kT$ determined from equation \eqref{eq:kT}. The amplitudes $A_i$ are scaled to agree with the relative contribution to the spectrum of each ion. \smallskip}
\label{tab:2}
\begin{ruledtabular}
\begin{tabular}{crrrrrrrr}
\multicolumn{1}{c}{Ion} & \multicolumn{1}{c}{$kT$} (eV) & \multicolumn{1}{c}{$A_1$} & \multicolumn{1}{c}{$\lambda_{c,1}$ (nm)} & \multicolumn{1}{c}{$w_1$ (nm)} &\rule[3.65mm]{0pt}{0mm}\rule[-1.8mm]{0pt}{0mm} & \multicolumn{1}{c}{$A_2$} & \multicolumn{1}{c}{$\lambda_{c,2}$ (nm)} & \multicolumn{1}{c}{$w_2$ (nm)} \\ \cline{3-5} \cline{7-9}
Sn$^{8+}$  & 17.8 & \rule[4.65mm]{0pt}{0mm}0.115 & 11.438 & 0.748 & & 0.247 & 11.438 & 0.094 \\
Sn$^{9+}$  & 21.7 & 0.346                                & 10.400  & 0.105 & & 0.244 & 10.400   & 0.560  \\
Sn$^{10+}$ & 25.9 & 0.267                                & 9.321  & 0.496 & & 0.763 & 9.705  & 0.127 \\
Sn$^{11+}$ & 30.3 & 1.000                                & 9.155  & 0.136 & & 0.641 & 8.647  & 0.158 \\
Sn$^{12+}$ & 35.1 & 0.700                                & 8.100   & 0.086 & & 0.775 & 8.650  & 0.127 \\
Sn$^{13+}$ & 40.1 & 0.408                                & 7.709  & 0.052 & & 0.426 & 8.210   & 0.085 \\
Sn$^{14+}$ & 45.3 & 0.123                                & 7.313  & 0.019 & & 0.244 & 7.740  & 0.019 \\
Sn$^{15+}$ & 50.9 & 0.234                                & 6.950   & 0.078 & & 0.628 & 7.380   & 0.055
\end{tabular}
\end{ruledtabular}
\end{table*} 
\newpage
\mbox{}
\newpage
\mbox{}
\section*{References}
\bibliographystyle{apsrev4-1}
\bibliography{library_v25}
\end{document}